\documentclass[prl,twocolumn,showpacs]{revtex4}

\usepackage{graphicx}
\usepackage{dcolumn}
\usepackage{bm}

\begin{document}

\title{Observation of a nonmonotonic transverse voltage induced by vortex motion in a
superconducting thin film}

\author{I.~Yu.~Antonova, V.~M.~Zakosarenko}
\author{E.~V.~Il'ichev}
\altaffiliation [ Now at ] { Institute for Physical High
Technology, P.O. Box 100239, D-07702 Jena, Germany}
\author{V.~I.~Kuznetsov}
\email{kvi@ipmt-hpm.ac.ru}
\author{V.~A.~Tulin}
\affiliation{Institute of Microelectronics Technology and High Purity Materials,
Russian Academy of Sciences, 142432 Chernogolovka, Moskow Region, Russia}

\begin{abstract}
A vortex-induced transverse voltage in a superconducting film,
previously predicted theoretically, has been seen experimentally
for the first time. The magnitude of this voltage and its
nonmonotonic current dependence are explained on the basis of a
curvature of the trajectory of vortices resulting from their
interaction. A good agreement between theory and experiment is
found. 
\end{abstract}

\maketitle

An interaction between magnetic flux vortices of opposite signs in a current-carrying
superconducting thin film can result in the appearance of a voltage in the direction transverse
with respect to the transport current. This effect was discussed theoretically by Glazman
\cite{Glazman}.
In this letter we are reporting an experimental confirmation of this effect.

Glazman's theory can be outlined as follows: The magnetic field of the transport current
penetrates into a strip of film in the form of vortices of opposite sign, because of the
different field directions at the opposite edges of the film. The attraction of the vortices
to each other causes their trajectories to become curved if they enter the film from the
opposite edges in regions displaced along the current. The result is the appearance of a
transverse voltage. The transverse current-voltage characteristic is not monotonic.

 \begin{figure}

 \includegraphics [width=1.0 \linewidth]{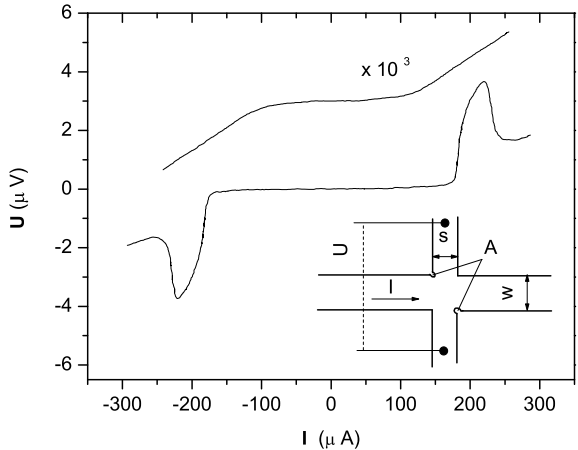}

\caption{\label{fig 1.} Longitudinal (upper curve) and transverse current-voltage
characteristics. The longitudinal characteristic has been displaced upward
for clarity, and the corresponding voltage scale is in millivolts. The inset
shows the central part of the sample.}

\end{figure}

The sample preparation procedure consisted of the following basic steps:
(1) vacuum deposition of tin to a thickness $d = 400 - 700$ {\AA} on silicon substrates;
(2) the formation of a pattern in the resist by electron lithography;
(3) etching of the film in an argon plasma. The inset in Fig. 1 shows the central part of a
sample. The sample is a film strip of width $w = 4 - 8$ \textit{$\mu$}m with transverse potential contacts.
The penetration of flux vortices into the film is facilitated by some notches $A$ displaced
along the current direction by the width of the potential contacts, $s = 3 - 5$ \textit{$\mu$}m.
The size of these notches is $ 0.2 - 0.7$ \textit{$\mu$}m.

 Longitudinal and transverse dc current-voltage characteristics are recorded in a
 superconducting shield, cooled in such a manner than the geomagnetic field is expelled from
 the shield when it goes superconducting. A magnetic field perpendicular to the film is
 produced by a copper solenoid inside the superconducting shield. The temperature is
 determined from the helium vapor pressure and is regulated by a membrance evacuation regulator.
 The derivative of the I-V characteristic is measured by means of a sonic-frequency modulation
 of the current.

 Figure 1 shows some typical I-V characteristics, longitudinal (the upper
curve) and transverse, recorded in a zero magnetic field. On the transverse
curve we see the voltage surges predicted in \cite{Glazman}. There are some differences
between the experimental and theoretical curves. First, the voltage peaks
arise at currents well above the $I_{c}$ of the film, while according to \cite{Glazman}
these peaks should appear at current near $I_{c}$. The other regions of the
film apparently have a lower critical current. The reason may lie in an
inhomogeneity of the film or in a current spreading in the crossing region,
since the width of the potential contacts is on the order of the film width.
Second, it was predicted in \cite{Glazman} that the voltage would drop to zero after a
surge on the theoretical transverse current-voltage characteristic. We see a
different picture: The voltage drops to a certain level and then exhibits a
linear behavior with increasing $I$. We believe that the reason for this
discrepancy is the nonaxial arrangement of the potential contacts, due in
particular to the presence of notches $A$. To test this interpretation, we
measured a transverse I-V characteristic with the sample in its normal
state, at $T > T_{c}$.
The curve turned out to be linear, with a resistance $ \approx 10^{-2}$ $\Omega$
for this particular structure. This result
corresponds to a $0.25-\mu$m misalignment of the potential contacts.

\begin{figure}

\includegraphics [width=1.0 \linewidth]{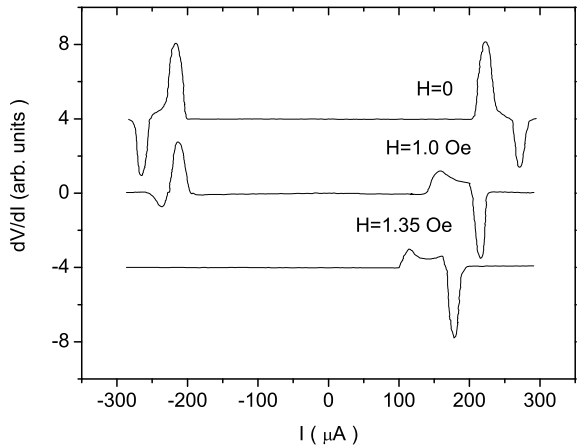}
\caption{\label{fig2:} Derivative of the transverse current-voltage characteristic recorded
at $T = 3.89$ K in various magnetic fields. The curves have been
displaced vertically for clarity.}

\end{figure}
  We can work from the experimental data to estimate the size of the voltage
surge which we would expect on the transverse current-voltage characteristic
as the result of the motion and annihilation of a vortex-antivortex pair.
The curves in Fig. 1 were found for a structure with $w = 7$ $\mu$m,
$s = 4$ $\mu$m, $d = 600$ {\AA} ,
$R_{square} = 0.31$ $\Omega$, and $T_{c} = 3.94$ K at $T = 3.89$ K.
We use the expression from \cite{Glazman}, which takes into account the displacement of
the vortex entry points, $s$:

\begin{equation} V  \simeq \hbar \Phi_{0}^{2}/8 \pi^{2} we \eta d \lambda_{\perp}s.
 \end{equation}

 \noindent Here $e$ is the charge of an electron, $\Phi_{0} = hc/2e$ is the flux quantum, and
$\lambda_{\perp} = 2\lambda^{2}/d$, where $\lambda$ is the magnetic-field penetration depth.
The viscosity $\eta$ can be estimated from \cite{Bardeen} $\eta \simeq \Phi_{0} H_{c2} /
\rho c^{2}$. Using $H_{c2} = \Phi_{0} /2 \pi \xi^{2}$ and $\rho / d = R_{square}$, we then find
the product

\begin{equation}
 \eta d \simeq \Phi_{0}^{2} / 2 \pi c^{2} \xi^{2} R_{square}.
 \end{equation}

 \noindent The electron mean free path $l$ can be estimated from the resistance at
liquid-helium temperature under the assumption \cite{Mermin} $ \rho$ $l  \simeq 1.6 \times
10^{-11}$ $\Omega cm^{2}$. This estimate yields $l \simeq 900$ {\AA}. Since the coherence
length for tin is  $ \xi_{0} = 2300$ {\AA}  \cite{Duzer}, the subsequent calculations are
carried out for the case of a dirty superconductor: \cite{Gorkov}

\begin{equation} \xi(T) = 0.85( \xi_{0} l)^{1/2} (1 - T/T_{c})^{-1/2}.
\end{equation}

\

\noindent The penetration depth can be estimated from the value of $R_{square}$ \cite{Likharev}:

\begin{equation} \lambda _{\perp} ( \mu m) \simeq 0.83 R_{square} (\Omega m) / (T_{c} - T).
 \end{equation}

\noindent The calculations yield $\lambda_{\perp} \simeq 5 $ $\mu $m and $\eta d \simeq 2 \times
10^{-15}$ cgs units. We thus find $V \simeq 1$ $\mu $V.
These rather simple estimates agree well with the experimental results (Fig. 1).
Equating the Lorentz force to the friction force (ignoring
pinning), we can estimate the vortex velocity: $\upsilon  = \Phi_{0}I / cwd \eta \simeq 1
\times 10^{6}$ cm/s. This value is on the order of the sound velocity in the material.

Figure 2 shows the derivative of the transverse voltage, \textit{dU/dI}, versus $I$, according
to measurements in various magnetic fields. At $H = 0$ the curve is symmetric.
With increasing field, an asymmetry with respect to the direction of the
transport current arises, and in a field $H = 2.3$ Oe the structural features in
the transverse voltage disappear completely. We believe that this asymmetry
occurs because notches $A$ are not exactly equivalent.

In summary, the effect predicted by Glazman \cite{Glazman} has been observed, and there
is a quantitative agreement between theory and experiment.

We wish to thank V. T. Volkov and A. N. Pronin for etching the structures.


\begin{references}

\bibitem{Glazman}
L.~I.~Glazman, Fiz. Nizk. Temp. {\bf 12(7)}, 688 (1986) [Sov. J. Low Temp. Phys. {\bf 12},
389 (1986)].

\bibitem{Bardeen}
I.~Bardeen and M.~I.~Stephen, Phys. Rev. {\bf 140}, 1197 (1965).

\bibitem{Mermin}
N.~W.~Ashcroft and N.~D.~Mermin, \textit{Solid State Physics}, Holt, Rinehart, and Winston,
New York, 1976.

\bibitem{Duzer}
T.~Van~Duzer and O.~Turner (eds.), \textit{Principles of Superconductivity Devices and
Circuits}, Elsevier, New York, 1981.

\bibitem{Gorkov}
L.~P.~Gor'kov, Zh. Eksp. Teor. Fiz. {\bf 36}, 1918 (1959) [Sov. Phys. JETP 9, 1364 (1959)].

\bibitem{Likharev}
K.~K.~Likharev, Zh. Eksp. Teor. Fiz. {\bf 61}, 1700 (1971) [Sov.Phys. JETP {\bf 34},
906 (1971)].



\end{references}
\end{document}